\documentclass[aps,twocolumn,showpacs,preprintnumbers,floatfix,nofootinbib,floatrow]{revtex4-1}
\usepackage{graphicx}
\usepackage{subcaption}
\usepackage{epsfig}
\usepackage{epstopdf}
\usepackage{dcolumn}
\usepackage{amsmath}
\usepackage{lipsum}

\def\be{\begin{equation}}
\def\ee{\end{equation}}
\def\bea{\begin{eqnarray}}
\def\eea{\end{eqnarray}}

\usepackage{color}      

\begin{document}

\title{\vspace{1.0in} {\bf The nucleon isovector tensor charge from lattice QCD using chiral fermions}}

\author{Derek Horkel$^{1}$, Yujiang Bi$^{2}$, Martha Constantinou$^{1}$, Terrence Draper$^{3}$, Jian Liang$^{3}$, Keh-Fei Liu$^{3}$, Zhaofeng Liu$^{2,4}$, Yi-Bo Yang$^{5,6}$
\vspace*{-0.5cm}
\begin{center}
\large{
\vspace*{0.4cm}
\includegraphics[scale=0.20]{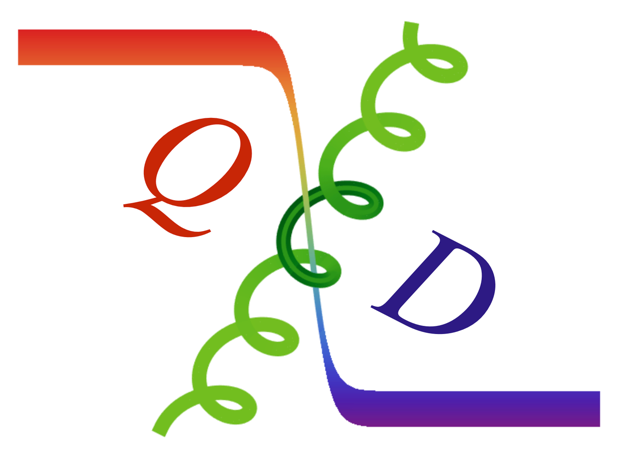}\\
\vspace*{0.4cm}
($\chi$QCD Collaboration)
\vspace*{0.75cm}
}
\end{center}
}
\affiliation{
$^{1}$\mbox{Department of Physics, Temple University, Philadelphia, PA 19122-1801, USA}\\ \vskip -0.2cm
$^{2}$\mbox{Institute of High Energy Physics, Chinese Academy of Sciences, Beijing 100049, China}\\  \vskip -0.2cm
$^{3}$\mbox{Department of Physics and Astronomy, University of Kentucky, Lexington, KY 40506, USA}\\ \vskip -0.2cm
$^{4}$\mbox{School of Physics, University of Chinese Academy of Sciences, Beijing 100049, China}\\ \vskip -0.2cm
$^{5}$\mbox{Department of Physics and Astronomy, Michigan State University, East Lansing, MI 48824, USA}\\ \vskip -0.2cm
$^{6}$\mbox{CAS Key Laboratory of Theoretical Physics, Institute of Theoretical Physics, }\\
\mbox{Chinese Academy of Sciences, Beijing 100190, China}}

\begin{abstract}

\vspace*{1cm}
In this work we present the isovector flavor combination for the nucleon tensor charge extracted from lattice QCD simulations using overlap fermions on $N_f=2+1$ domain-wall configurations. The pion mass dependence is studied using six valence quark masses, each reproducing a value for the pion mass in the valence sector between 147 and 330 MeV. We investigate and eliminate systematic uncertainties due to contamination by excited states, by employing several values for the source-sink separation that span from 1 fm to 1.6 fm. We apply a chiral extrapolation in the valence sector using a quadratic and a logarithmic term to fit the pion mass dependence, which describes well the lattice data. The lattice matrix element is renormalized non-perturbatively, and the final result is $g_T=1.096(30)$ in the $\overline{\rm MS}$ scheme at a renormalization scale of 2 GeV.

\end{abstract}

 \maketitle

\section{Introduction}
\label{sec:intro}

Parton distribution functions (PDFs) are important tools to understand the structure of hadrons and have played an important role in establishing QCD as the theory of the strong interaction. These quantities are universal, and therefore, experimental data from different processes are analyzed together within the global analysis framework for the extraction of the $x$-dependent PDFs. There are several collaborations (e.g., ABMP~\cite{Alekhin:2017kpj}, CJ~\cite{Accardi:2016qay}, CT~\cite{Dulat:2015mca}, HERAPDF~\cite{Abramowicz:2015mha}, JR~\cite{Jimenez-Delgado:2014twa},  MMHT~\cite{Harland-Lang:2014zoa},  NNPDF~\cite{Ball:2017nwa}) analyzing the available experimental data which have obtained satisfactory agreement for the chiral-even PDFs, that is, the unpolarized ($f_1$) and helicity ($g_1$) PDFs. On the contrary, the collinear transversity PDF ($h_1$) is poorly known, as a limited number of experimental data are available, only in certain kinematic regions, and are less precise.

The PDFs cannot be calculated directly on a Euclidean lattice, as they correspond to non-local operators with time-like separated fields. Instead, it is more straightforward to calculate their Mellin Moments, leading to matrix elements of a tower of local operators, which are well-studied. It is worth mentioning that novel pioneering approaches have been proposed for a more direct access to PDFs, the hadronic tensor~\cite{Liu:1993cv,Liu:1999ak,Liang:2019frk}, quasi-distributions~\cite{Ji:2013dva,Ji:2014gla}, pseudo-distributions~\cite{Radyushkin:2016hsy,Radyushkin:2017ffo,Radyushkin:2017cyf}, good lattice cross sections~\cite{Ma:2014jla,Ma:2017pxb}, to name a few. All of these methods are under investigation within lattice QCD and are summarized in the recent review of Ref.~\cite{Cichy:2018mum}. Of particular interest is the work of Ref.~\cite{Alexandrou:2018eet} which is the first complete calculation of the $x$-dependence of the transversity PDFs for the nucleon, with simulations at the physical point.

The tensor charge is the first Mellin Moment of the transversity PDF, and is a fundamental quantity in understanding the internal structure of hadrons. It is also related to physics beyond the Standard Model (BSM)~\cite{Dubbers:2011ns,DelNobile:2013sia,Bhattacharya:2011qm,Courtoy:2015haa} as, together with the scalar charge, it probes novel scalar and tensor interactions at the TeV scale. For example, neutron $\beta$-decay experiments require input on the scalar and tensor charges to provide reliable estimates. The nucleon tensor charge plays an important role also in searches of a nonzero electric dipole moment that originates from CP-violating contributions~\cite{Bhattacharya:2016zcn}.

While the nucleon tensor charge is not accurately known from the analysis of experiments and from phenomenology, it can be obtained from three-point functions in  lattice QCD with statistical and systematic uncertainties better controlled than by taking the moments from the lattice PDFs at this stage~\cite{Cichy:2018mum,Alexandrou:2018eet} . This gives a unique opportunity to combine lattice data with experimental measurements for a better constraint on the tensor charge. This has been tested in the analysis of Ref.~\cite{Lin:2017stx}, demonstrating that such synergy is realistic and promising.

Given the importance of the tensor charge, there is an on-going effort to better constrain its value. For example, there is a rich experimental program in the 12 GeV upgrade at Jefferson Lab, to investigate the transverse spin nucleon structure~\cite{Ye:2016prn,Dudek:2012vr}. New experiments in Hall A will employ a future solenoid spectrometer (SoLID) to perform precision measurements from semi-inclusive electro-production of charged pions from  transversely polarized $^3$He target in Deep-Inelastic-Scattering kinematics using 11 and 8.8 GeV electron beams~\cite{Gao:2010av}. SoLID is expected to increase the experimental accuracy of the tensor coupling by an order of magnitude~\cite{Dudek:2012vr,Ye:2016prn}. Also, current experiments at LHC are probing scalar and tensor interactions for BSM physics at the TeV scale. The transversity PDFs at large $x$ are also included in the physics program of the future Electron-Ion-Collider~\cite{Accardi:2012qut,Aschenauer:2014twa}, endorsed by the National Academies of Science, Engineering and Medicine~\cite{NAP25171}. Thus, a precise determination of the tensor charge from lattice QCD is crucial and timely.

Recent years have seen marked progress in lattice QCD mainly due to algorithmic advances and an increase of computational resources. The synergy of the above has enabled simulations to be carried out at parameters close to or at their physical values, and high-accuracy results are now available for the tensor charge (see, e.g, Refs.~\cite{Bhattacharya:2016zcn,Hasan:2019noy,Alexandrou:2019brg}). This has enabled an intense activity within lattice QCD to provide high-precision input to experiments, test phenomenological models, and predict physics beyond the Standard Model.

\smallskip
This paper is organized as follows: The methodology and numerical implementation are explained in Section~\ref{sec:num}, while the renormalization procedure is laid out in Section~\ref{sec:renorm}. In Section~\ref{sec:res}, we show our results for the tensor charge and a detailed discussion of the control of the excited-states contamination and the chiral extrapolation. Finally, in Section~\ref{sec:sum} we summarize and discuss our final results.

\section{Numerical details}
\label{sec:num}

\begin{figure}[htb]
  \includegraphics[scale=0.5,angle=0]{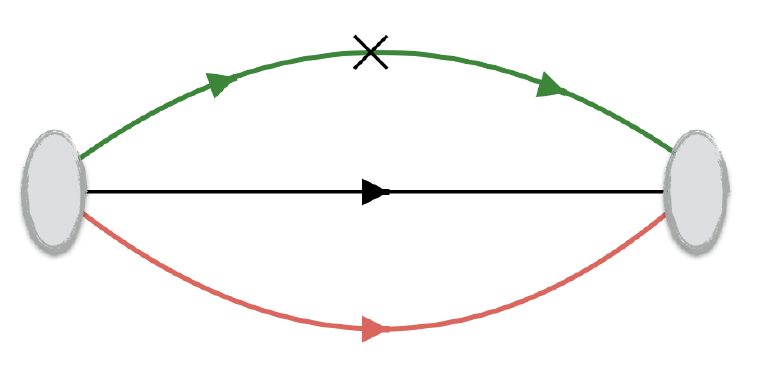}
  \captionsetup{justification=raggedright}
\caption{Example of the connected insertion.} \label{fig:ci}
\end{figure}

The isovector tensor charge can be accurately computed from lattice QCD, as it is extracted directly from lattice data. It receives contributions only from the connected insertion as in Fig.~\ref{fig:ci}, in which the tensor operator is inserted to the quark propagator on the connected insertion.

In this work, we use a single ensemble of $N_f=2+1$ RBC/UKQCD domain-wall fermions (DWF) Iwasaki gauge configurations~\cite{Blum:2014tka}, and the overlap formulation in the valence sector. The lattice spacing corresponding to the gauge configurations (32ID) is $0.1431(7)$ fm and has a volume of $32^3\times 64$~\cite{Arthur:2012yc}. Thus, the spatial extent of the lattice in physical units is approximately 4.6 fm, and allows one to reach the near-physical sea pion mass of 170 MeV.

An advantage of overlap fermions is the fact that one can generate the quark propagators with multiple quark masses at a small additional computational cost compared to the cost of the lightest quark mass. We employ six values for the quark masses that give a valence pion mass ($m_\pi$) ranging from 147 MeV to 327 MeV. Therefore, one can perform a partially quenched chiral extrapolation to the physical point in the valence sector. Another advantage of the overlap formulation is that the effective overlap operator ($D_c$) we use is chiral, i.e. $\{D_c, \gamma_5\} = 0$ \cite{Chiu:1998gp}. $D_c$ is expressed in terms of the overlap operator $D_{ov}$ as
\bea
D_c=\frac{\rho D_{ov}}{1-\frac{D_{ov}}{2}} \textrm{ with }D_{ov}=1+\gamma_5\epsilon(\gamma_5D_{\rm w}(\rho)),
\eea
where $\epsilon$ is the matrix sign function and $D_{\rm w}$ is the Wilson Dirac operator with a negative mass
characterized by the parameter $\rho=4-1/2\kappa$ for $\kappa_c < \kappa < 0.25$. In this work we set $\kappa$=0.2 which corresponds to $\rho=1.5$. Details on the calculations with overlap fermion can be found in our previous work~\cite{Li:2010pw}.

The matrix elements we need are obtained from the ratio of the three-point function to the two-point function
\begin{equation}   \label{eq:ratio}
 \hspace*{-0.25cm} R(t,T_{\rm sink})=\frac{\hspace*{-0.2cm}\langle 0|{\int} d^3 y\Gamma_k^m{\chi}(\vec{y},t_f)\bar{\psi}(t)\frac{\sigma_k}{3}\psi(t){\displaystyle{\sum_{\vec{x}\in G}}}\bar{\chi}_S(\vec{x},0)|0 \rangle}{\langle 0|\int d^3 y\Gamma^e\chi(\vec{y},t_f)\sum_{\vec{x}\in G}\bar{\chi}_S(\vec{x},0)|0 \rangle},
\end{equation}
\vskip 0.1cm
\noindent where $\chi$ is the standard proton interpolation field and $\bar{\chi}_S$ is the grid $Z_3$ noise source with gaussian smearing applied to all three quarks with size $\sim 0.6$ fm. The source is located at timeslice $t_0=0$, the sink at $t_f$ and the current insertion, $t$, varies between the source and the sink. We indicate the time separation between the source and the sink as $T_{\rm sink}$. The two-point functions are projected with the unpolarized projector,
\vspace*{-0.5cm}
\begin{equation}
\Gamma^e=\frac{1}{2}(1+\gamma_4)\,,
\end{equation}
\vskip -0.5cm
\noindent while the three-point functions require a polarized projector,
 \vspace*{-0.5cm}
 \begin{equation}
 \Gamma^m_k=\frac{i}{2}(1+\gamma_4)\gamma_k\gamma_5\,,
 \end{equation}
\vskip -0.5cm
\noindent to extract $g_T$. The polarized projector is along the spatial direction  $k$, while the current is defined as $\sigma_k=\epsilon_{ijk}\gamma_i\gamma_j$. We obtain the three-point functions for each value of $k$, and then we sum over the spatial directions $k=1,2,3$ with the appropriate current insertion that gives nonzero signal. All the correlation functions from the source points $\vec{x}$ in the grid $G$ are combined to improve the signal-to-noise ratio.

To reliably extract the tensor charge, the source-sink separation $T_{\rm sink}$ has to be large enough to suppress excited-states contamination. Also, the insertion time $t$ is taken to be away from the source and the sink, to guarantee ground state dominance in the ratio $R(t,T_{\rm sink})$. To examine effects from excited states we compute $R(t,T_{\rm sink})$ for five values of the sink-source separation $T_{\rm sink}/a=7, 8, 9, 10, 11$, corresponding to 1.00, 1.14, 1.29, 1.43, 1.57 fm, respectively. For each $T_{\rm sink}$ value, the current is inserted at all timeslices from the source to the sink.

In this work, we use the stochastic method with low mode substitution (LMS) to generate the two-point and three-point functions efficiently. We use six 2-2-2-2 grid sources, and 2/4/6/10/12 stochastic wall source for the cases with $T_{\rm sink}/a=7, 8, 9, 10, 11$ respectively to approximate the corresponding sequential sources, on 200 configurations. The projection $(1-\gamma_4)/2$ is applied to the backward nucleon propagators and thus the total measurements are 6(sources)*16(points in the grid)*2(forward and backward)*200(configurations)=38,400. Note that we have the same statistics for each valence quark mass, as the overlap inverter can be applied to multiple masses without much overhead. To suppress the additional statistical uncertainty from the stochastic sources, the LMS is applied to all four quark propagators. The details of the simulation setup can be found in Refs.~\cite{Gong:2013vja,Yang:2015zja,Liang:2016fgy}.

\section{Renormalization}
\label{sec:renorm}

One important ingredient for extracting the tensor charge is the renormalization of the operator that removes divergences related to the regulator, as well as the leading dependence on the fermionic and gluonic action (up to cut-off effects). The bare matrix elements of the tensor charge are renormalized multiplicatively, and we compute the appropriate renormalization function, $Z_T$, non-perturbatively in the RI/MOM scheme. This is then converted to the $\overline{\textrm MS}$ scheme using continuum perturbation theory~\cite{Gracey:2000am,Gracey:2003yr} and evolved to a scale of 2 GeV that allows comparison with phenomenological estimates. Below we explain the procedure followed to obtain the value used in this work.

Based on the strategy presented in our previous paper~\cite{Liu:2013yxz} we first determine the renormalization function of the local axial-vector current, $Z^{\rm WI}_A$, using the Ward Identity (WI), that is
\bea
Z^{\rm WI}_A=\frac{2m_q\langle 0| \bar{\psi}\gamma_5\psi |\pi\rangle}{m_{\pi}\langle 0| \bar{\psi}\gamma_5\gamma_4 \psi|\pi\rangle}.
\eea
One may use $Z^{\rm WI}_A$ to extract the RI/MOM renormalization function of the tensor charge, which is defined as
\bea
Z_T^{\rm RI/MOM}(\mu)=Z^{\rm WI}_A\left.
\frac{\textrm{Tr}[(\gamma_k\gamma_5)^{-1}\Lambda_{\gamma_k\gamma_5}(p)]}{\textrm{Tr}[\sigma^{-1}_k\Lambda_{\sigma_k}(p)]}\right|_{p^2=\mu^2}.\label{eq:renorm_gt}
\eea
In the above equation, $\Lambda_{\Gamma}(p)$ is the forward vertex function
\bea
\Lambda_{\Gamma}(p)=S^{-1}(p)\sum_{x,y} e^{-ip\cdot(x-y)}\langle \psi(x) {\cal O}(0) \bar\psi(y)\rangle S^{-1}(p)\qquad
\eea
with $S(p)=\sum_x e^{-ip\cdot x}\langle \psi(x) \bar\psi(0)\rangle$ being the quark propagator in momentum space, and ${\cal O}$ the tensor operator. Such a definition does not require knowledge of the  quark field renormalization, extracted from the quark self energy, which may have large discretization errors~\cite{Liu:2013yxz}. In our previous study we have demonstrated a similar procedure for the renormalization function of the vector current, which was found to be consistent with the one determined from the vector charge~\cite{Yang:2015zja}.

$Z_T^{\rm RI/MOM}$ is gauge dependent and the vertex functions and quark propagators are computed in the Landau gauge. We also employ periodic boundary conditions in all four directions, and the momentum $p$, which is set to the RI/MOM renormalization scale $\mu$, is chosen as
\begin{equation}
ap=2\pi\left(\frac{k_1}{L}, \frac{k_2}{L}, \frac{k_3}{L}, \frac{k_4}{T}\right)\,.
\end{equation}
The integer appearing in the components of the momentum is chosen as $k_\mu=-8,-7,...,8$, and the lattice size is $L=32$ and $T=64$. To reduce the effects of Lorentz
non-invariant discretization errors, we apply a ``democratic" cut~\cite{Constantinou:2010gr}
\begin{equation}
\frac{p^{[4]}}{(p^2)^2}{<}0.32,\quad\mbox{where } p^{[4]}=\sum_\mu p_\mu^4,\quad p^2=\sum_\mu p_\mu^2\,.
\label{eq:p_cut}
\end{equation}

\begin{figure}[htb]
  \includegraphics[width=1.0\hsize,angle=0]{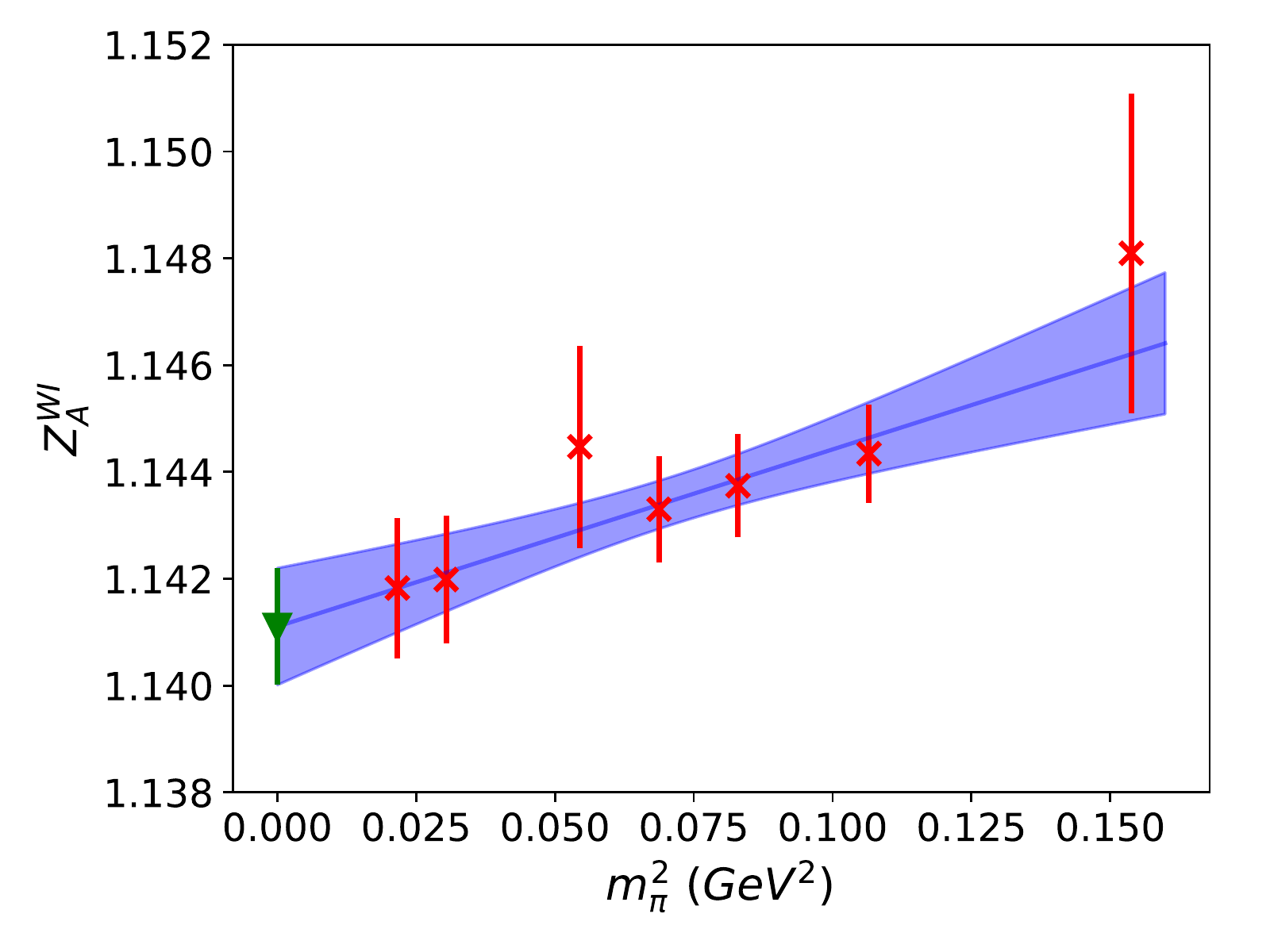}
  \captionsetup{justification=raggedright}
\caption{The renormalization function of the axial-vector operator from the Ward Identity, as a function of the pion mass dependence (red points). The green triangle shows the chirally extrapolated value.} \label{fig:renorm_1}
\end{figure}
The extracted value for $Z^{\rm WI}_A$ and its dependence on the valence pion mass is plotted in Fig.~\ref{fig:renorm_1}. The errors shown are statistical uncertainties computed using the jackknife method and are of the order of $\sim$0.1\% using 200 configurations. As can be seen from the plot, the pion mass dependence is very weak and within statistical uncertainties. The difference in $Z_A$ between the lowest and the heaviest value of $m_\pi$ is less than $\sim$0.3\%. We perform a chiral extrapolation using a linear fit in $m_\pi^2$, and the chiral value of $1.141(1)$ is shown with a green triangle. The number in the parenthesis corresponds to the statistical uncertainty.

\begin{figure}[htb]
\hspace*{-1cm}  \includegraphics[scale=1,angle=0]{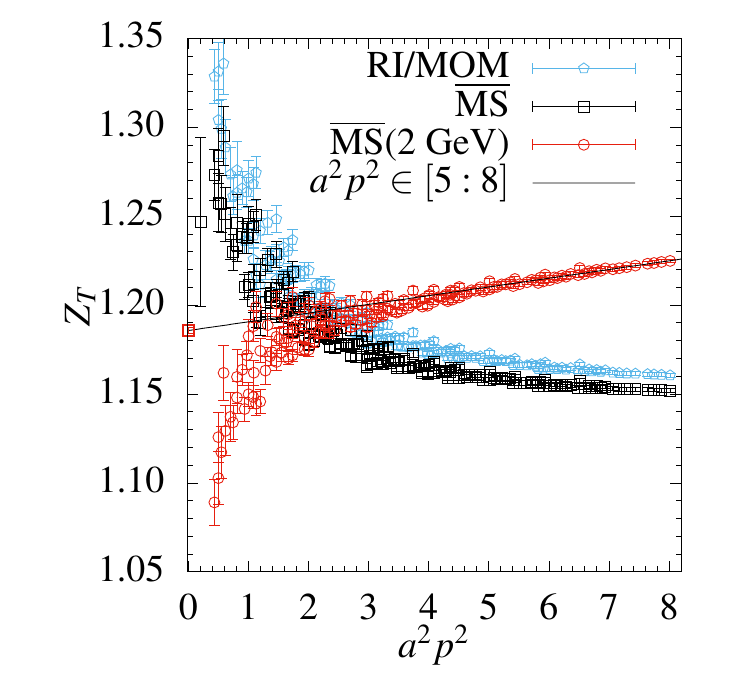}
\captionsetup{justification=raggedright}
\caption{$Z_T$ as a function of the initial RI/MOM scale $(ap)^2$. RI/MOM values are shown with light blue polygons, and $\overline{\textrm{MS}}$ values at a scale $(ap)^2$ are shown with black squares. Red circles correspond to $\overline{\textrm{MS}}$ values evolved to $\mu=2$ GeV. The final value used in this work is obtained from the $(ap)^2 {\to} 0$ limit using a linear extrapolation in $(ap)^2$, which is applied on the red points.}
 \label{fig:renorm_2}
\end{figure}

Fig.~\ref{fig:renorm_2} shows the renormalization function of the tensor operator in the RI/MOM scheme (blue polygons) as defined in Eq.~(\ref{eq:renorm_gt}). The statistical uncertainties of $Z_T^{\rm RI/MOM}$ at $(ap)^2{>}3$ are less than 0.5\%, which is the region of interest for the final fit. A chiral extrapolation in the valence quark sector has already been applied to the data in the plot, similarly to the case of $Z_A$.
The black squares correspond to $Z_T$ upon conversion to the $\overline{\textrm{MS}}$ scheme at the same scale as in the initial RI/MOM scheme, that is, $\mu^2=p^2$. Finally, the red circles show $Z_T$ in the $\overline{\textrm{MS}}$ scheme  evolved at a common scale for all points, $\mu=2$ GeV. The conversion and evolution use 4-loop expressions for the anomalous dimension in $\overline{\textrm{MS}}$ extracted in continuum perturbation theory~\cite{Gracey:2000am,Gracey:2003yr}. The $\overline{\textrm{MS}}$ estimates at a fixed scale are expected to have a constant behavior up to ${\cal O}((ap)^2)$ effects, which are found to be non-negligible as the initial scale $(a p)^2$ increases. To remove the residual dependence on the initial RI/MOM scale we perform a  linear extrapolation in $(ap)^2{\to}0$ obtaining a value of 1.1857(17). For the aforementioned fit we use the range $(ap)^2\in[5,8]$ and the corresponding $\chi^2/$d.o.f is $0.7$.

 In this work we examined the following systematic uncertainties on the renormalization function, and considered the ones contributing above 0.1$\%$: \\[1ex]
{\bf{1.}} Truncation effects in the conversion factor to the $\overline{\textrm{MS}}$ scheme and evolution of scale. This is estimated by comparing the results using 3-loop and 4-loop formulas. Above $(ap)^2=5$, this error is about or less than 0.1\%; \\[0.75ex]
{\bf{2.}} Uncertainty in $\Lambda_{\rm QCD}^{\overline{\textrm{MS}}}=339(10)$ MeV for evaluating $\alpha_s$. Varying $\Lambda_{\rm QCD}^{\overline{\textrm{MS}}}$ from 339 MeV to 349 MeV changes the central value of $Z_T^{\overline{\textrm{MS}}}(2\mbox{ GeV})$ by 0.14\%; \\[0.75ex]
{\bf{3.}} Uncertainty in the value used for the lattice spacing ($1/a=1.3784(68)$GeV) when choosing $(ap)^2$ such that $\mu=2$ GeV. This effect is found to be less than $0.1\%$ and thus negligible; \\[0.75ex]
{\bf{4.}} Variation of the final value with the fit range for obtaining $(ap)^2 {\to} 0$. We vary $(ap)^2$ from [5,8] to [4,8], which leads to a 0.23\% change in
$Z_T^{\overline{\textrm{MS}}}(2\mbox{ GeV})$. \\[1ex]
The uncertainties due to $\Lambda_{\rm QCD}^{\overline{\textrm{MS}}}$ and the fit range are added quadratically to get the total systematic error. Thus, we report as our final result:
\begin{equation}
Z_T^{\overline{\textrm{MS}}}(2\mbox{ GeV})=1.1857(17)(36)\,.
\end{equation}
The aforementioned systematic effects are found to be similar to the renormalization functions of other operators, such as, the scalar operator~\cite{Liu:2013yxz}.

\section{Results}
\label{sec:res}

In this section we present the analysis for excited-states contamination, with main focus on the plateau method and the two-state fits. The first method relies on a constant fit in a region where a plateau is identified, that is
\begin{equation}
R(t,T_{\rm sink})_{\overrightarrow{{T_{\rm sink}-t\rightarrow \infty} \atop {t-t_i\rightarrow \infty}}} \Pi(T_{\rm sink})\,.
\label{eqn:Rplateau}
\end{equation}
In the above equation, $t_i$ is the insertion time of the source, which is zero in our case. The tensor charge is obtained upon renormalization,
\begin{equation}
g_T^{\rm plateau}(T_{\rm sink}) = Z_T  \Pi(T_{\rm sink})\,.
\label{eqn:Rplateau}
\end{equation}
No additional kinematic factor is needed due to the use of the rest frame.

An alternative analysis approach for the isolation of the ground state is to perform a two-state fit, assuming dominance of the ground and first excited states. In such a method, the ratio is fitted to the form,
\vspace*{-.35cm}
\begin{equation}
R(t,T_{\rm sink}) = R_0 + C_1\, e^{-\delta m\, (T_{\rm sink}-t)} + C_2\, e^{-\delta m\, t}\,,
\label{eqn:RTfit}
\end{equation}
for each quark mass. The tensor charge is then given by
\begin{equation}
g_T^{\rm 2\mbox{-}state} = Z_T  R_0\,.
\end{equation}
 In principle, there is an additional term, $C_3\, e^{-\delta m T_{\rm sink}}$, in this expression. However, it was found to be insignificant in the fit, and thus $C_3$ was set to zero for better stability in the fit. Once the ground state contribution is obtained successfully, we expect that $g_T^{\rm 2\mbox{-}state}$ will be consistent with the value extracted from the plateau fit at some large source-sink separation.

\begin{widetext}
\begin{figure*}[ht!]
\centerline{(a) $m_\pi=0.3266$ GeV  \hspace*{4cm} (b) $m_\pi=0.2881$ GeV  \hspace*{1cm} }
   \begin{subfigure}[t]{0.5\textwidth}
      \centering
\hspace*{-3cm} \includegraphics[scale=0.61,angle=0]{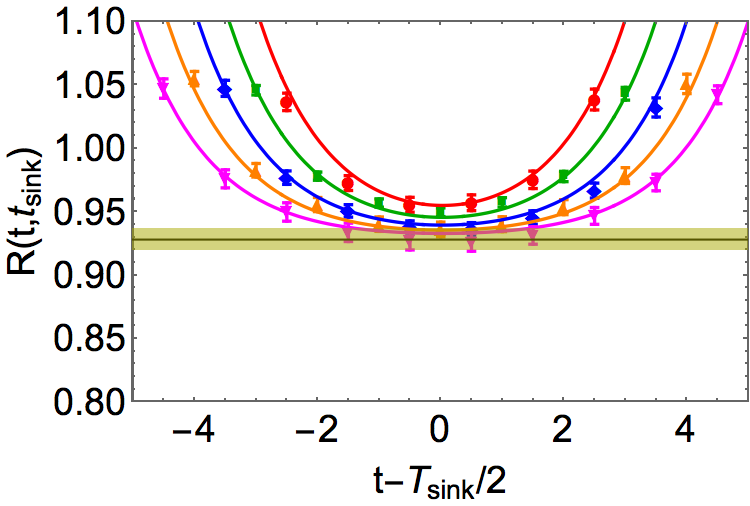}
      \label{fig:Mass22000}
   \end{subfigure}
   \begin{subfigure}[t]{0.4\textwidth}
      \centering
\hspace*{-1.5cm}  \includegraphics[scale=0.61,angle=0]{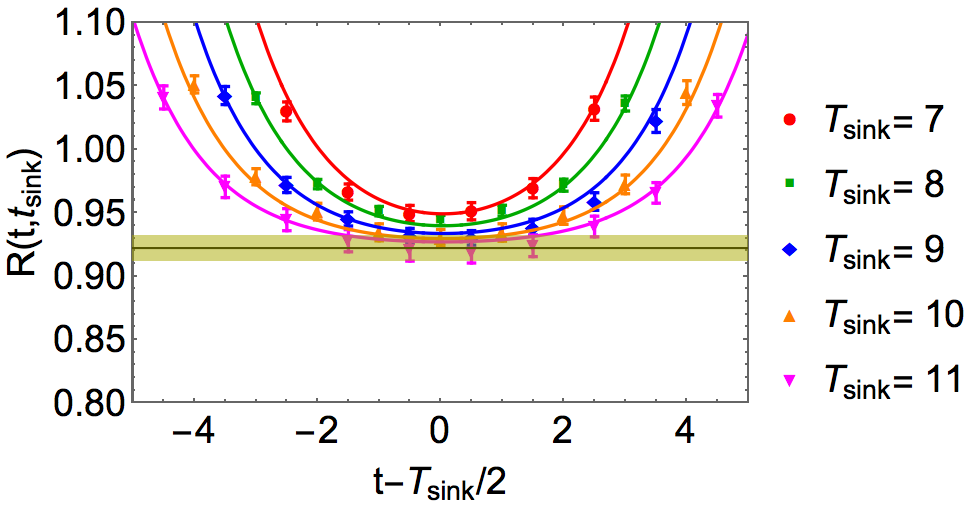}
      \label{fig:Mass17000}
   \end{subfigure}

\centerline{(c) $m_\pi=0.2623$ GeV  \hspace*{4cm} (d) $m_\pi=0.2335$ GeV  \hspace*{1cm} }
   \begin{subfigure}[t]{0.5\textwidth}
      \centering
\hspace*{-3cm} \includegraphics[scale=0.61,angle=0]{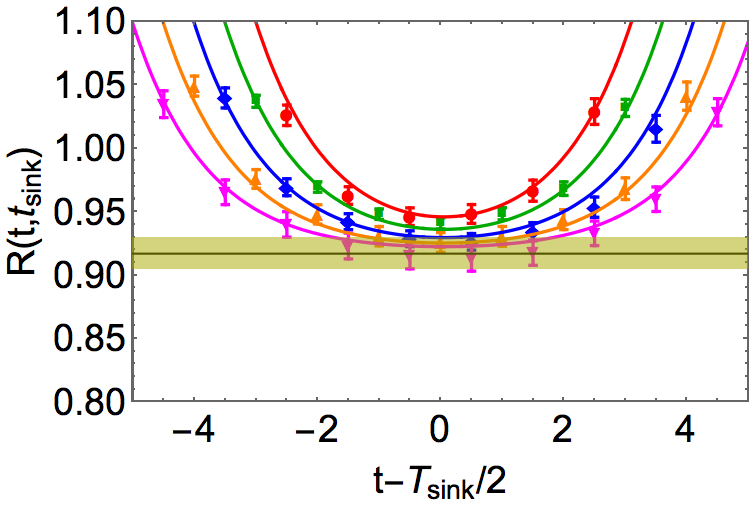}
      \label{fig:Mass14000}
   \end{subfigure}
   \begin{subfigure}[t]{0.4\textwidth}
      \centering
\hspace*{-1.5cm}  \includegraphics[scale=0.61,angle=0]{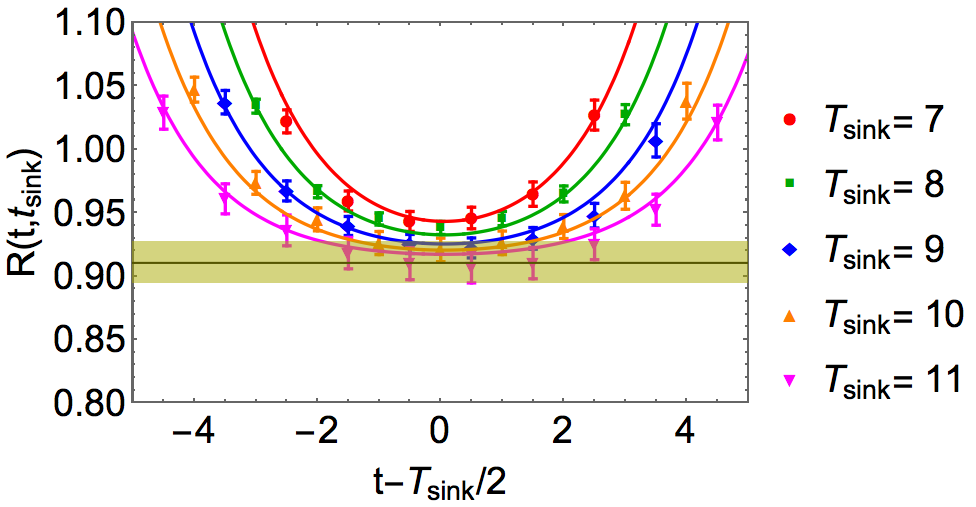}
      \label{fig:Mass11000}
   \end{subfigure}

\centerline{(e) $m_\pi=0.1745$ GeV  \hspace*{4cm} (f) $m_\pi=0.1471$ GeV  \hspace*{1cm} }
   \begin{subfigure}[t]{0.5\textwidth}
      \centering
\hspace*{-3cm} \includegraphics[scale=0.61,angle=0]{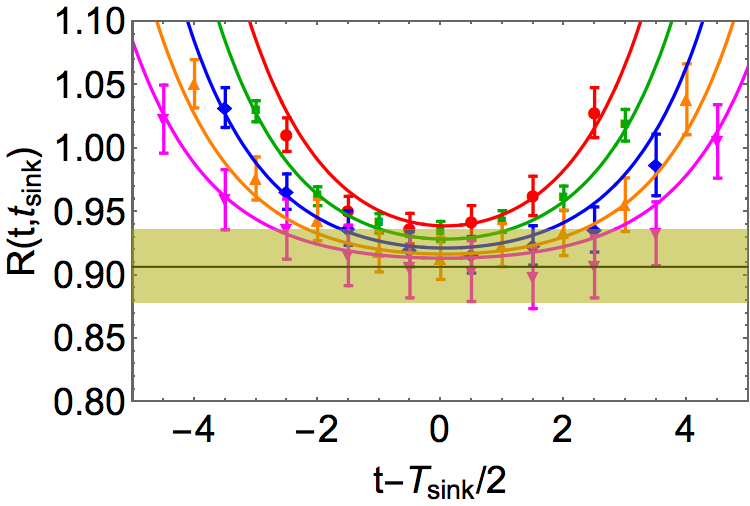}
      \label{fig:Mass6000}
   \end{subfigure}
   \begin{subfigure}[t]{0.4\textwidth}
      \centering
\hspace*{-1.5cm}  \includegraphics[scale=0.61,angle=0]{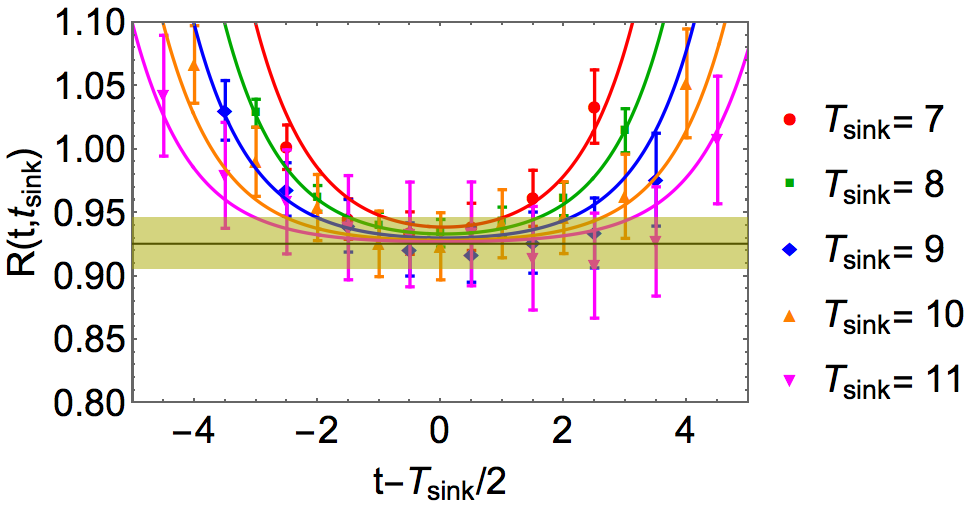}
      \label{fig:Mass4200}
   \end{subfigure}
   \captionsetup{justification=raggedright}
   \caption{$R(t,T_{\rm sink})$ as a function of the insertion time $t$ for all six ensembles. The two-state fit of Eq.~(\ref{eqn:RTfit}) is shown for each value of $T_{\rm sink}$: 7$a$ (red), 8$a$ (green), 9$a$ (blue), 10$a$ (orange), 11$a$ (magenta). The extracted value of $g^B_T=R_0$ is shown with a band.}
      \label{fig:R_Mass}
\end{figure*}
\end{widetext}

The unrenormalized ratio $R(t,T_{\rm sink})$ of Eq.~(\ref{eq:ratio}) is plotted in Fig.~\ref{fig:R_Mass} for each quark mass used in this work. The ratio for source-sink separation $7a,\,8a,\,9a,\,10a,\,11a$ is shown with red, green, blue, orange and magenta points, respectively. With a black constant line we show $R_0$ extracted from the two-state fit of Eq.~(\ref{eqn:RTfit}), and its width (green band) shows the statistical uncertainties. For demonstration purposes, we keep the range of the $y$-axis the same for each plot so a direct comparison between different quark masses can be made. Starting from the top left plot and moving to the bottom right plot, we show the results for $m_\pi=326.6$, 288.1, 262.3, 233.5, 174.5, and 147.1 MeV. From the analysis of the various quark masses we find that excited-states contamination is very small. Comparing the results from different values of the pion mass gives us a comprehensive understanding of the dependence of excited states as the pion mass decreases. As can be seen from the plots, the two-state fit is compatible with the one-state fit using data at $T_{\rm sink}>1.1$ fm. This is in agreement with other investigations on the same quantity (see, e.g., Ref.~\cite{Alexandrou:2019brg}). For the quark masses with $m_\pi<200$ MeV we find that the excited-states effect is within the statistical uncertainties for all separations. The individual plateau values for $T_{\rm sink}=7a,\,9a,\,11a$ are given in Table~\ref{tab:gTcompare}, and are compared to the values extracted from the two-state fit.

\begin{table}[h]
\begin{center}
\begin{tabular}{ c|c|c|c|c }
\,$m_\pi$ (GeV)\, & \,$T_{\rm sink}=7a$\, &  \,$T_{\rm sink}=9a$\, &  \,$T_{\rm sink}=11a$\, &  \,two-state\,  \\
    \hline\hline
0.3266            &1.144(07)    &1.117(06)    &1.103(08)     & 1.100(10)  \\[0.5ex]
0.2881            &1.136(08)    &1.110(07)    &1.094(11)     & 1.093(12)  \\[0.5ex]
0.2623            &1.132(08)    &1.106(08)    &1.087(12)     & 1.087(15)  \\[0.5ex]
0.2335            &1.129(09)    &1.102(09)    &1.080(14)     & 1.079(19)  \\[0.5ex]
0.1745            &1.122(15)    &1.097(17)    &1.074(28)     & 1.075(34)  \\[0.5ex]
0.1471            &1.118(23)    &1.098(26)    &1.103(49)     & 1.097(24)  \\[0.5ex]
    \hline
  \end{tabular}
  \end{center}
  \vspace*{-0.25cm}
       \captionsetup{justification=raggedright}
  \caption{Renormalized $g_T$ for each value of the pion mass using the plateau method of Eq.~(\ref{eqn:Rplateau}) ($T_{\rm sink}=7a,\,9a,\,11a$) and the two-state fit given in  Eq.~(\ref{eqn:RTfit}).}
\label{tab:gTcompare}
  \end{table}

\begin{figure}[h!]
  \includegraphics[width=1.0\hsize,angle=0]{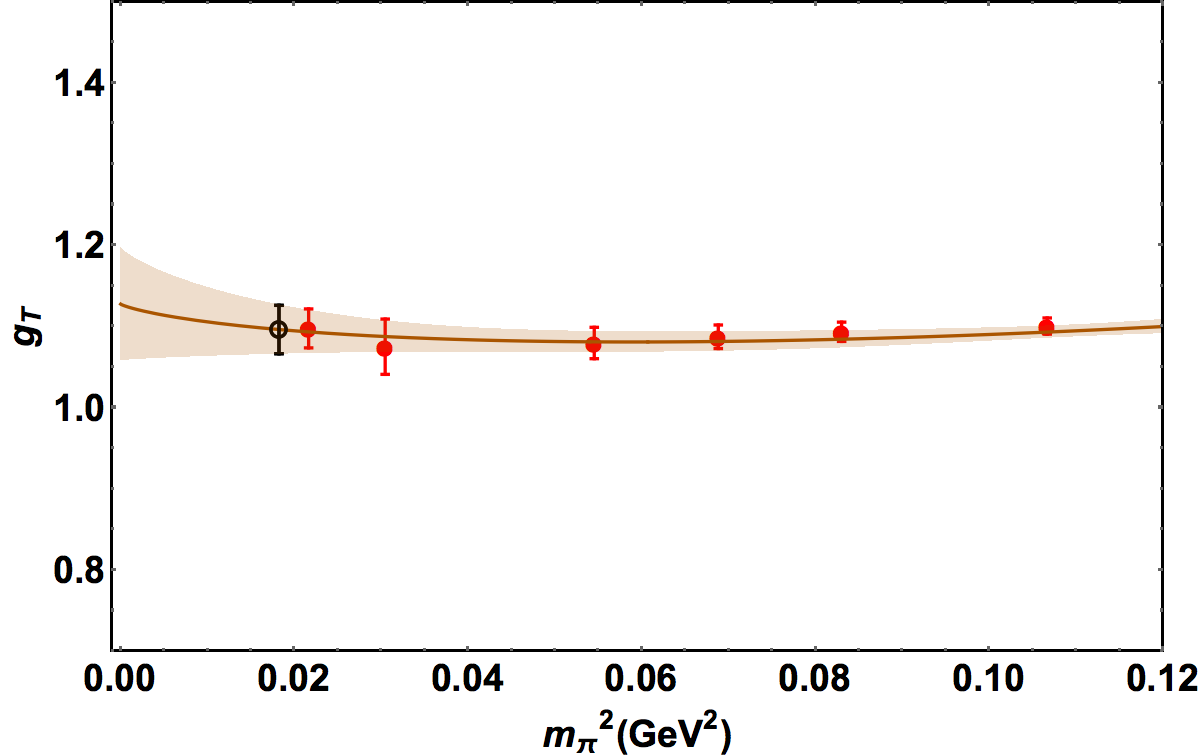}
     \captionsetup{justification=raggedright}
   \caption{$g_T$ as a function of the pion mass squared (red points), together with the chiral extrapolation of Eq.~(\ref{eqn:chiral}) (red band). Each point was obtained using the two-state fit method. The empty black point is the extrapolated value. The error for each ensemble is calculated using the jackknife method. The error on the extrapolated point and error band are calculated using the super-jackknife method.}
   \label{chiral_extrap}
\end{figure}

   \vspace*{0.25cm}
In order to obtain results at the physical pion mass, we use a chiral extrapolation with respect to the valence pion mass. We fit the data to the form
\begin{equation}
g_T = a + b~ m_\pi^2 + c~ m_\pi^2 \log(\frac{m_\pi^2}{m_\rho^2})\,,
\label{eqn:chiral}
\end{equation}
where $m_\rho=0.775$ GeV. The final extrapolated value at the physical point is
\begin{equation}
g_T= 1.096(30)\,,
\end{equation}
obtained using the two-state fit values at each quark mass. The statistical error has been determined using the super-jackknife method~\cite{Bratt:2010jn}. The final results on each ensemble are shown with red points in Fig.~\ref{chiral_extrap}, while the fit function of Eq.~(\ref{eqn:chiral}) is shown with a red band. We observe a rather flat behavior with respect to the pion mass. In fact, omitting the last term in Eq.~(\ref{eqn:chiral}) gives compatible results with the fit shown in the plot. The extrapolated value is shown with a black open circle and is obtained from a fit including the logarithmic term.

It is interesting to compare our final result with the recent work of Refs.~\cite{Bhattacharya:2016zcn,Hasan:2019noy,Alexandrou:2019brg} on the isovector combination for the tensor charge. While the results from these references correspond to simulations directly at the physical point, a comparison is justified by the mild dependence on the pion mass. In addition, such a comparison can give an indication of the effectiveness  of the chiral fit. A calculation by PNDME~\cite{Bhattacharya:2016zcn} uses 9 ensembles of $N_f=2+1+1$  HISQ fermions at different lattice spacings, volumes and pion mass ($\sim130 - 320$ MeV). This allows for a combined continuum, chiral and infinite-volume extrapolation, obtaining as a final estimate $g_T=0.987(51)$. LHPC performed a calculation on $N_f=2+1+1$ of 2-HEX-smeared Wilson-clover fermions, using two ensembles at the physical pion mass and different values of the lattice spacing~\cite{Hasan:2019noy}. They find $g_T=0.972(24)$ for $a=0.093$ fm and $g_T=0.989(23)$ for $a=0.116$ fm. ETMC has recently obtained $g_T$ on three ensembles at the physical point using $N_f=2$ and $N_f=2+1+1$ twisted mass fermions at volumes up to 5 fm~\cite{Alexandrou:2019brg}. The reported results are $g_T=0.992(22)$ ($N_f=2$, $L=4.5$ fm) $g_T=0.974(33)$ ($N_f=2$, $L=6$ fm) $g_T=0.926(32)$ ($N_f=2+1+1$, $L=5.1$ fm). Tension between our result and results from other formulations, indicates further systematic uncertainties (e.g., finite lattice spacing, volume effects) not yet addressed. Given that the continuum limit has not been taken, comparison between the various groups is only qualitative.

\section{Summary}
\label{sec:sum}

We have presented a lattice calculation of the nucleon tensor charge using a mixed setup of chiral fermions, that is, overlap fermions on $N_f=2+1$ domain-wall configurations. Lattice results for this quantity are very important, as they may be used to constrain global fits of the transversity PDFs, due to lack of experimental data in all kinematic regions. Besides the tensor charge's significance in hadronic physics, it is also related to physics beyond the Standard Model. We focus on the isovector combination which has no contributions from disconnected diagrams. The valence pion mass ranges between 147 and 330 MeV, while the pion mass in the sea sector of the RBC/UKQCD action is 170 MeV, which is close to its physical value. At each quark mass we study excited-states contamination using several values of the source-sink time separations, between 1 fm to 1.6 fm. These data allow us to perform one- and two-state fits for the elimination of excited-states contamination, as well as a chiral extrapolation to physical quark masses in the valence sector. The final result after chiral extrapolation is $g_T=1.096(30)$ and is given in the $\overline{\rm MS}$ scheme at a scale of 2 GeV. It is worth mentioning that we find a very weak pion mass dependence, and the uncertainty of the chiral extrapolation is comparable to the statistical error of $g_T$ at the lightest valence pion mass. In the near future we intend to address further systematic uncertainties, by including more ensembles of gauge configurations at different values of the lattice spacing and volume.

\section*{Acknowledgments}

The work presented in this paper received financial support by the U.S. Department of Energy, Office of Nuclear Physics, within
the framework of the TMD Topical Collaboration. The project is supported in part by the U.S. DOE Grant No.\ DE-SC0013065. M.C. is partly supported by the National Science Foundation under Grant No.\ PHY-1714407.
Z.L. acknowledges the support of the National Science Foundation of China
under Grants No.\ 11575197.
Y.Y. is partly supported by the Chinese Academy of Science CAS Pioneer  Hundred Talents Program. This research used resources of the Oak Ridge Leadership Computing Facility at the Oak Ridge National Laboratory, which is supported by the Office of Science of the U.S. Department of Energy under Contract No.\ DE-AC05-00OR22725. Part of the data were generated using resources from the Extreme Science and Engineering Discovery Environment (XSEDE), which is supported by National Science Foundation grant number ACI-1053575. We also thank the National Energy Research Scientific Computing Center (NERSC) for providing HPC resources that have contributed to the research results reported within this paper. This work also supported by the Strategic Priority Research Program of Chinese Academy of Sciences, Grant No.\ XDC01040100. 

\bibliographystyle{apsrev4-1}
\bibliography{reference.bib}

\end{document}